\documentclass[12pt]{article}

\usepackage{graphicx}

\begin{document}

\begin{titlepage}
\begin{flushright}

NTUA--02--2002

hep-th/0202044 \\

\end{flushright}

\begin{centering}
\vspace{.41in}
{\large {\bf Brane Cosmology}}\\

\vspace{.2in}

 {\bf E.~Papantonopoulos}$^{a}$  \\
\vspace{.2in}

 National Technical University of Athens, Physics
Department, Zografou Campus, GR 157 80, Athens, Greece. \\

\vspace{1.0in}
\begin{abstract}
\vspace{.2in}

The aim of these lectures is to give a brief introduction to brane
cosmology. After introducing some basic geometrical notions, we
discuss the cosmology of a brane universe with matter localized on
the brane. Then we introduce an intrinsic curvature scalar term in
the bulk action, and analyze the cosmology of this induced
gravity. Finally we present the cosmology of a moving brane in the
background of other branes, and as a particular example, we
discuss the cosmological evolution of a test brane moving in a
background of a Type-0 string theory.
\end{abstract}
\end{centering}

\vspace{1.3in}
\begin{flushleft}

Lectures presented at the First Aegean Summer School on Cosmology,
Samos, September 2001.

 \vspace{.2in} $^{a}$ e-mail address:lpapa@central.ntua.gr \\

\end{flushleft}

\end{titlepage}

\section{Introduction}

Cosmology today is an active field of physical thought and of
exiting experimental results. Its main goal is to describe the
evolution of our universe from some initial time to its present
form. One of its outstanding successes is the precise and detailed
description of the very early stages of the universe evolution.
Various experimental results confirmed that inflation describes
accurately these early stages of the evolution. Cosmology can also
help to understand the large scale structure of our universe as it
is viewed today. It can provide convincing arguments why our
universe is accelerating and it can explain the anisotropies of
the Cosmic Microwave Background data.

The mathematical description of Cosmology is provided by the
Einstein equations. A basic ingredient of all cosmological models
is the matter content of the theory. Matter enters Einstein
equations through the energy momentum tensor. The form of the
energy momentum tensor depends on the underlying theory. If the
underlying theory is a Gauge Theory, the scalar sector of the
theory must be specified and in particular its scalar potential.
Nevertheless, most of the successful inflationary models, which
rely on a scalar potential, are not the result of an uderlying
Gauge Theory, but rather the scalar content is arbitrary fixed by
hand.

In String Theory the Einstein equations are part of the theory but
the theory itself is consistent only in higher than four
dimensions. Then the cosmological evolution of our universe is
studied using the effective four dimensional String Theory. In
this theory the only scalar available is the dilaton field. The
dilaton field appears only through its kinetic term while a
dilaton potential is not allowed. To have a dilaton potential with
all its cosmological advantages, we must consider $a^{\prime}$
corrections to the String Theory. Because of this String Theory is
very restrictive to its cosmological applications.

The introduction of branes into cosmology offered another novel
approach to our understanding of the Universe and of its
evolution. It was proposed \cite{pap.Reg} that our observable
universe is a three dimensional surface (domain wall \index{domain
wall}, brane \index{brane}) embedded in a higher dimensional
space. In an earlier
 speculation, motivated by the long standing hierarchy problem, it
 was proposed \cite{pap.Dim} that the fundamental Planck scale could
 be close to the gauge unification scale, at the price of "large"
 spatial dimensions, the introduction of which explains the
 observed weakness of gravity at long distances. In a similar
 scenario \cite{pap.Rand}, our observed world is embedded in a
 five-dimensional bulk, which is strongly curved. This allows the
 extra dimension not to be very large, and we can perceive
 gravity as effectively four-dimensional.

\begin{figure}[h]
\centering
\includegraphics[scale=0.6]{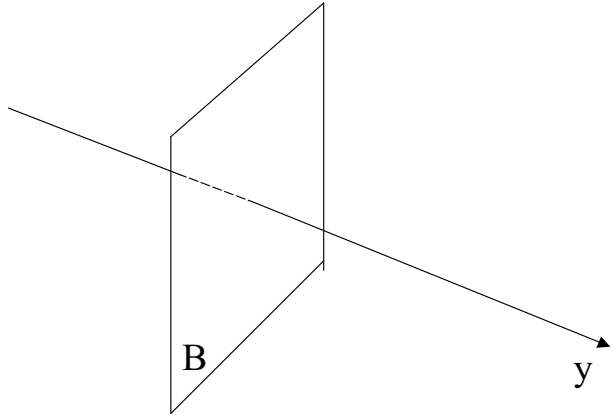}\caption
{A brane embedded in a five dimensional space.}
\end{figure}

 This idea of a brane universe \index{brane?brane universe}
 can  naturally be applied to String
 Theory. In this context, the Standard Model gauge bosons as well as
 charged matter arise as fluctuations of the D-branes. The universe
 is living on a collection of coincident branes, while gravity and
 other universal interactions is living in the bulk space
 \cite{pap.Pol}.

 This new perception of our world had opened new directions in
 cosmology, but at the same time imposed some new problems. The
 cosmological evolution of our universe should take place on the
 brane, but for the whole theory to make sense, the brane should
 be embedded in a consistent way to a higher dimensional space
 the bulk. The only physical field in the bulk is the
 gravitational field, and there are no matter fields. Nevertheless
 the bulk leaves its imprint on the brane, influencing in this way
 the cosmological evolution of our universe. In the very early
 attempts to study the brane cosmology, one of the main problems
 was, how to get the standard cosmology on the brane.

\begin{figure}[h]
\centering
\includegraphics[scale=0.5]{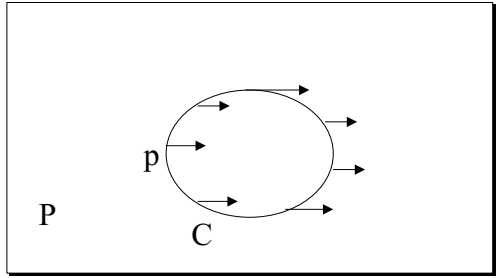}\caption
{A vector moving around a curve $C$.}
\end{figure}

 In these lectures,  which are addressed to the first years
  graduates students, we will discuss the cosmological evolution
  of our universe on the brane. Our approach would be more
  pedagogical, trying to provide the basic
 ideas of this new cosmological set-up, and more importantly to
 discuss the mathematical tools necessary for the construction of
 a brane cosmological model. For this reason will not exhaust all
 the aspects of brane cosmology. For us, to simplify things, a
 brane \index{brane}
 is a three dimensional surface which is
 embedded in a higher dimensional space which has only one extra
 dimension, parametrized by the coordinate $y$, as it is shown in
 Fig.1.

 The lectures are organized as follows. In section two after an
 elementary geometrical description of the extrinsic curvature, we will
 describe in some detail the way we embed a D-dimensional surface
 in a D+1-dimensional bulk. We believe that understanding this
 procedure is crucial for being able to construct a brane
 cosmological model. In section three we will present the Einstein
 equations on the brane and we will solve them for  matter
 localized on the brane. Then we will discuss the Friedmann-like
 equation we get on the brane and the ways we can recover the
 standard cosmology. In section four we will see what kind of
 cosmology we get if we introduce in the bulk action a
 four-dimensional curvature scalar. In section five we will
 consider a brane moving in the gravitational field of other
 branes, and we will discuss the cosmological evolution of a test
 brane moving in the background of a type-0 string theory. Finally
 in the last section we will summarize the basic ideas and results of brane
 cosmology.

\section{A Surface $\Sigma$ embedded in a D-dimensional Manifold
$M$}

 \subsection{Elementary Geometry}

To understand the procedure of the embedding of a surface $\Sigma$
in a higher dimensional manifold, we need the notion of the
extrinsic curvature \index{curvature?extrinsic curvature}. We
start by explaining the notion of the curvature \cite{pap.wald}.

 If you have a plane $P$ and a curve $C$ on it, then any
vector starting from the point $p$ and moving along the curve $C$
it will come at the same point p with the same direction as it
started as shown in Fig.2. However, if the surface was a three
dimensional sphere, and a vector
 starting from $p$ is moving along $C$, will come back on
$p$, having different direction from the direction it started
with, as shown in Fig.3.

These two examples give the notion of the curvature
\index{curvature}. The two dimensional surface is flat, while the
three dimensional surface is curved. Having in mind Fig.2 and
Fig.3 we can say that a space is curved if and only if some
initially parallel geodesics \index{geodesic} fail to remain
parallel. We remind to the reader that a geodesic is a curve whose
tangent is parallel-transported along itself, that is a
"straightest possible" curve.

\begin{figure}[h]
\centering
\includegraphics[scale=0.45]{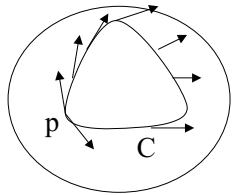}\caption
{A vector moving around a curve $C$ on a sphere.}
\end{figure}

We can define the notion of the parallel transport of a vector
along a curve $C$ with a tangent vector $t^{a}$, using the
derivative operator $\nabla_{a}$. A vector $v^{a}$ given at each
point on the curve is said to be parallely transported, as one
moves along the curve, if the equation
\begin{equation}\label{pap.parral}
t^{a}\nabla_{a}v^{b}
\end{equation}
is satisfied along the curve. Consider a point $q$ on $\Sigma$
with a normal vector $n^{a}$. If I parallel transport the vector
$n^{a}$ to a point $p$, then it will be the dashed lines. The
failure of this vector to coincide with the vector $n^{a}$ at $p$
corresponds intuitively to the bending of $\Sigma$ in the space
time in which is embedded (Fig.4). This is expressed by the
extrinsic curvature
\begin{equation}\label{pap.extrcurv}
K_{ab}=h^{c}_{a}\nabla_{c}n_{b}
\end{equation}
where $h_{ab}$ the metric on $\Sigma$.

\begin{figure}[h]
\centering
\includegraphics[scale=0.5]{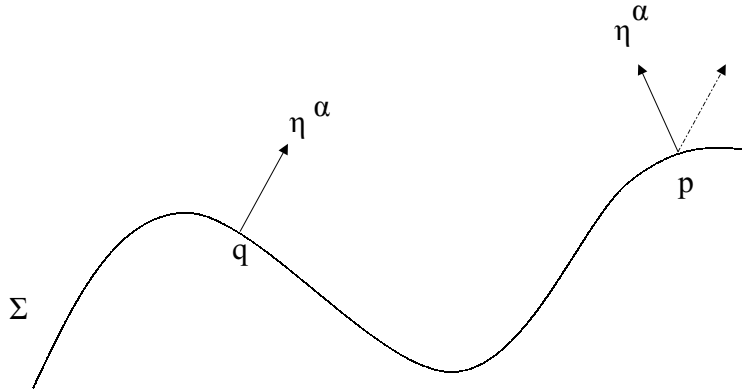}\caption
{The notion of extrinsic curvature.}
\end{figure}

\subsection{The Embedding Procedure}

Imagine now that we have a surface $\Sigma$ (Known also as a
domain wall or brane) embedded in a D-dimensional Manifold $M$
\cite{pap.Cha}. Assume that M splits in two parts $M_{\pm}$. We
demand the metric to be continuous everywhere and the derivatives
of the metric to be continuous everywhere except on $\Sigma$. The
Einstein-Hilbert action on $M$ is
\begin{equation}\label{pap.einaction}
S_{EH}=-\frac{1}{2}\int d^{D}x \sqrt{-g} R
\end{equation}
where $g_{MN}$ is the metric on $M$, with M,N=1,...,D. We define
the induced metric \index{induced metric} on $\Sigma$ as
\begin{equation}
h_{MN}=g_{MN}-n_{M}n_{N}
\end{equation}
where $n_{M}$ is the unit normal vector into $M\pm$. We vary the
action (\ref{pap.einaction}) in $M\pm$ and we get
\begin{equation}\label{pap.var1}
\delta S_{EH}=-\frac{1}{2} \int_{\Sigma \pm} d^{D-1} \sqrt{-h}
g^{MN}n^{p}(\nabla_{M} \delta g_{NP}-\nabla_{P} \delta g_{MN})
\end{equation}
If we replace $g^{MN}=h^{MN}+n^{M}n^{N}$ in (\ref{pap.var1}) we
get
\begin{equation}\label{pap.var2}
\delta S_{EH}=-\frac{1}{2} \int_{\Sigma \pm} d^{D-1} \sqrt{-h}
h^{MN}n^{p}(\nabla_{M} \delta g_{NP}-\nabla_{P} \delta g_{MN})
\end{equation}
Recognize the term $(\nabla_{M} \delta g_{NP}-\nabla_{P} \delta
g_{MN})$ as the discontinuity of the derivative of the metric
across the surface. We do not like discontinuities so we introduce
on both side of the surface $\Sigma$, the Gibbons-Hawking
\index{Gibbons-Hawking term} boundary term
\begin{equation}\label{pap.hawgib}
 S_{GH}=-\frac{1}{2} \int_{\Sigma \pm} d^{D-1} \sqrt{-h} K
\end{equation}
where $k=h^{MN}K_{MN}$ and $K_{MN}$ is the extrinsic curvature
defined in (\ref{pap.extrcurv}). If we vary the Gibbons-Hawking
term we get as expected
\begin{equation}\label{pap.var3}
\delta S_{GH}=-\frac{1}{2} \int_{\Sigma \pm} d^{D-1}
\sqrt{-h}(\delta K-\frac{1}{2} K h^{MN}\delta g_{MN})
\end{equation}
we need the variation of $\delta K$. If we use the variation
\begin{equation}
\delta n_{M}=\frac{1}{2}n_{M}n^{P}n^{Q}\delta g_{PQ}
\end{equation}
after some work we get
\begin{equation}\label{pap.var4}
\delta K= -K^{MN} \delta g - h^{MN}n^{P} (\nabla_{M} \delta
g_{NP}-\nabla_{P} \delta g_{MN}) +\frac{1}{2}K n^{P}n^{Q} \delta
g^{PQ}
\end{equation}
We have the variation of both terms $S_{EH}$ and $S_{GH}$ in
(\ref{pap.var2}) and (\ref{pap.var3}). Puting everything together
we get
\begin{eqnarray}\label{pap.whovar}
\delta S_{EH}+\delta S_{GH}&=&\int d^{D-1}x \sqrt{-h}
\Big{(}\frac{1}{2}h^{MN}n^{P}\nabla_{M} g_{NP}+K^{MN}\delta
g_{MN}\\ \nonumber &-&\frac{1}{2}K n^{M}n^{N} \delta g_{MN}-
\frac{1}{2} K h^{MN}\delta g_{MN}\Big{)}
\end{eqnarray}
To simplify the above formula imagine a vector $X^{M}$ tangential
to $\Sigma \pm$, then
\begin{equation}\label{pap.rel1}
\nabla_{M}X^{M}=h^{MN}\nabla_{M}X_{N}+n^{M}n^{N}\nabla_{M}X_{N}
\end{equation}
Define the derivative operator $\widetilde{\nabla}$ on $\Sigma$
from
\begin{equation}\label{pap.rel2}
\widetilde{\nabla}_{R}X^{M}=h^{M}_{P}h^{Q}_{R}X^{P}
\end{equation}
then, using ({\ref{pap.rel1}) and (\ref{pap.rel2}) we have
\begin{equation}
\nabla_{M}X^{M}=\widetilde{\nabla} _{M}X^{M}-X^{M}
n^{N}\nabla_{N}X_{M}
\end{equation}
and
\begin{equation}
h^{MN}n^{P}\nabla_{M}\delta g_{NP}=\nabla_{M}(h^{MN}n^{P}\delta
g_{NP})-\delta g_{NP} \nabla_{M}(h^{MN}n^{P})
\end{equation}
If we use the definition of extrinsic curvature $K_{MN}$ from
(\ref{pap.extrcurv}) we get
\begin{equation}\label{pap.rel3}
h^{MN}n^{P}\nabla_{M}\delta g_{NP}=\nabla_{M}(h^{MN}n^{P}\delta
g_{NP})+Kn^{M}n_{N} \delta g_{MN}-K^{MN} \delta g_{MN}
\end{equation}
If we substitute (\ref{pap.rel3}) in (\ref{pap.whovar}) and
integrate out the total derivative term we get our final result
\begin{equation}\label{pap.finalresul}
\delta S_{EH}+\delta S_{GH}=\frac{1}{2}\int_{\Sigma\pm} d^{D-1}x
\sqrt{-h}(K^{MN}-Kh^{MN}) \delta g_{MN}
\end{equation}
Therefore what we have done is that starting with the
Einstein-Hilbert action, we were forced to introduce the
Gibbons-Hawking term, to cancel the discontinuities and the
variation of both terms is expressed in terms of the extrinsic
curvature and its trace.

\subsection{The Israel Matching Conditions}

Relation (\ref{pap.finalresul}) is therefore the result of the
embedded surface $\Sigma$ into $M$, a pure geometrical process.
Now we can put some dynamics on the surface assuming that there is
matter on the surface with an action of the form
\begin{equation}\label{pap.lmat}
S_{M}=\int_{\Sigma\pm} d^{D-1}x \sqrt{-h}L_{matter}
\end{equation}
where $L_{matter}$ represents the matter on the brane. Then, the
variation of (\ref{pap.lmat}) gives
\begin{equation}\label{pap.lvar1}
\delta S_{M}=\int_{\Sigma\pm} d^{D-1}x \sqrt{-h}T^{MN}\delta
g_{MN}
\end{equation}
where $ T^{MN}\equiv -\frac{2}{\sqrt{-h}} \frac{\delta
S_{M}}{\delta h_{MN}}$ is the energy momentum tensor of the brane.
Now, if we demand the variation of the whole action
\begin{equation}
\delta S =\delta S_{EH}+\delta S_{GH}+\delta S_{M}
\end{equation}
to be zero, using (\ref{pap.finalresul}) and (\ref{pap.lvar1}) we
get the Israel Matching Conditions \index{Israel Matching
Conditions} \cite{pap.israel}
\begin{equation}\label{pap.isrmat}
\Big{\{} K_{MN}-K h_{MN}\Big{\}}=-T_{MN}
\end{equation}
where the curly brackets denote summation over both sides of
$\Sigma$. This relation is central for constructing any brane
cosmological model. In some way it supplements the Einstein
Equations in such a way as to make them consistent on the brane.

\section{Brane Cosmology \index{brane?brane cosmology}
 in 5-dimensional Spacetime}
\subsection{The Einstein Equations on the Brane}

The most general action describing a three-dimensional brane in a
five-dimensional spacetime is \cite{pap.kofinas}
\begin{eqnarray}\label{pap.action}
S_{(5)}&=&\frac{1}{2k^{2}_{5}}\int_{M} d^{5}x
\sqrt{-g}\Big{\{}^{(5)}R-2\Lambda_{5}\Big{\}}+\frac{1}{2k^{2}_{4}}\int_{\Sigma}
d^{4}x \sqrt{-g}\Big{\{}^{(4)}R-2\Lambda_{4}\Big{\}}\nonumber \\
&+&\int_{M} d^{5}x \sqrt{-g}L^{mat}_{5}+\int_{\Sigma} d^{4}x
\sqrt{-g}L^{mat}_{4}
\end{eqnarray}
where $\Lambda_{5}$ and $\Lambda_{4}$ are the cosmological
constants \index{cosmological constants} of the bulk and brane
respectively, and $L^{mat}_{5}$, $L^{mat}_{4}$ are their matter
content. From the dimensionful constants $k^{2}_{5}$, $k^{2}_{4}$
the Planck masses $M_{5}$, $M_{4}$ are defined as
\begin{eqnarray}\label{pap.const}
k^{2}_{5}&=&8\pi G_{(5)}=M^{-3}_{5} \nonumber \\ k^{2}_{4}&=&8\pi
G_{(4)}=M^{-2}_{4}
\end{eqnarray}
We will derive the Einstein equations  for the simplified action
\begin{equation}\label{pap.saction}
S_{(5)}=\frac{1}{2k^{2}_{(5)}}\int_{M} d^{5}x \sqrt{-g}^{(5)}R+
\int_{M} d^{5}x \sqrt{-g}L^{mat}_{5}
\end{equation}
which is historically the first action considered
\cite{pap.binutry}, and we will leave for later the general case.
We will consider a metric of the form
\begin{equation}
ds^{2}=g_{MN}dx^{M}dx^{N}=g_{\mu\nu}dx^{\mu}dx^{\nu}+b^{2}dy^{2}
\end{equation}
where $y$ paramertizes the fifth dimension. We will assume that
our four dimensional surface is sited at $y=0$. We allow a time
dependence of the fields so our metric becomes
\begin{equation}\label{pap.tmetric}
ds^{2}=-n^{2}(t,y)dt^{2}+a^{2}(t,y)\delta _{ij}
dx^{i}dx^{j}+b^{2}(t,y)dy^{2}
\end{equation}

Note that we take for simplicity flat metric for the ordinary
spatial dimensions. For the matter content of the action
(\ref{pap.saction}) we assume that matter is confined on both,
brane and bulk. Then, the energy momentum tensor derived from
(\ref{pap.saction}) can be decomposed into
\begin{equation}\label{pap.enmom}
T^{M}_{~~N}=T^{M}_{~~N}\Big{|}_{bulk}+T^{M}_{~~N}\Big{|}_{brane}
\end{equation}
For the matter on the brane we consider perfect fluid with
\begin{equation}\label{pap.rpcom}
T^{M}_{~~N}\Big{|}_{brane}=\frac{\delta(y)}{b}
diag(-\rho,p,p,p,,0)
\end{equation}
What we want now is to study the dynamics of the metric
$g_{\mu\nu}(t,0)$. For this, we have to solve the 5-dimensional
Einstein equations
\begin{equation}\label{pap.beinst}
G_{MN}=k^{2}_{5}T_{MN}
\end{equation}
Inserting (\ref{pap.tmetric}) in (\ref{pap.beinst}) we get
\begin{eqnarray}\label{pap.comp}
G_{00}&=&3\Big{\{} \frac{\dot{a}}{a}(\frac{\dot{a}}{a}
+\frac{\dot{b}}{b})-\frac{n^{2}}{b^{2}}\Big{(} \frac{a^{''}}
{a}+\frac{a^{'}}{a} (\frac{a^{'}}{a} -
\frac{b^{'}}{b})\Big{)}\Big{\}}\nonumber \\
 G_{ij}&=&\frac{a^{2}}{b^{2}}\delta_{ij}\Big{\{}\frac{a^{'}}{a}
 (\frac{a^{'}}{a}+2\frac{n^{'}}{n}) - \frac{b^{'}}{b} (
\frac{n^{'}}{n} + 2 \frac{a^{'}}{a}) + 2\frac{a^{''}}{a} +
\frac{n^{''}}{n}\Big{\}} \nonumber \\ &+& \frac{a^{''}}{n^{2}}
\delta_{ij} \Big{\{} \frac{\dot{a}}{a} ( -\frac{\dot{a}}{a} +2
\frac{\dot{n}}{n}) + \frac{\dot{b}}{b} ( -2\frac{\dot{a}}{a} +
\frac{\dot{n}}{n}) -2\frac{\ddot{a}}{a} - \frac{\ddot{b}}{b}
\Big{\}} \nonumber \\ G_{05} &=& 3 (\frac{n{'}}{n}
\frac{\dot{a}}{a} + \frac{a^{'}}{a} \frac{\dot{b}}{b} -
\frac{\dot{a}^{'}}{a} ) \nonumber \\ G_{55}&=& 3 \Big{\{}
\frac{a^{'}}{a} ( \frac{a^{'}}{a} + \frac{n^{'}}{n}) -
\frac{b^{2}}{n^{2}}\Big{(} \frac{\dot{a}}{a} (\frac{\dot{a}}{a} -
\frac{\dot{n}}{n}) + \frac{\ddot{a}}{a}\Big{)} \Big{\}}
\end{eqnarray}

We are looking for solutions of the Einstein equations
(\ref{pap.beinst}) near or in the vicinity of $y=0$. At the point
$y=0$, where the brane is situated, we must take under
consideration the Israel Boundary Conditions. We can use relations
(\ref{pap.isrmat}) to calculate them or follow an easier way
\cite{pap.binutry}. We require that the derivatives of the metric
with respect to $y$, to be discontinuous at $y=0$. This means that
in the second derivatives of the quantities $a$ and $n$ a
distributional term will appear which will have the form $[a^{'}]
\delta (y)$ or $ [n^{'}] \delta (y)$ with
\begin{equation}\label{pap.disc}
[a^{'}]  =  a^{'} (0^{+}) - a^{'}(0^{-})
\end{equation}
\begin{equation}\label{pap.disc1}
 [n ^{'}] =
n^{'}(0^{+}) - n^{'}(0^{-})
\end{equation}

We can calculate the quantities (\ref{pap.disc}) and
(\ref{pap.disc1}) using equations (\ref{pap.comp}) and the
energy-momentum tensor (\ref{pap.rpcom})
\begin{eqnarray}\label{pap.ret}
\frac{ [a^{'} ]}{a_{0}b_{0}}& =& -\frac{k^{2}_{(5)}}{3} \rho
\nonumber \\ \frac{ [n^{'} ]}{n_{0}b_{0}}& =&
\frac{k^{2}_{(5)}}{3} (3p+2\rho)
\end{eqnarray}
where $a_{0}=a(t,0)$ and $b_{0}=b(t,0)$ and we have set n(t,0)=1.
If we use a reflection symmetry $y\longrightarrow -y$ the $(55)$
component of the Einstein equations (\ref{pap.beinst}) with the
use of  (\ref{pap.rpcom}) and (\ref{pap.ret}) becomes
\begin{equation}\label{pap.cosm}
\frac{\dot{a}^{2}_{0}}{a^{2}_{0}} + \frac{\ddot{a}_{0}}{a_{0}}= -
\frac{k^{4}_{(5)}}{36} \rho(\rho+3p) -
k^{2}_{(5)}\frac{T_{55}}{3b^{2}_{0}}
\end{equation}
This is our cosmological Einstein equation which governs the
cosmological evolution of our brane universe \index{brane?brane
universe}.

\subsection{Cosmology on the Brane}

Define the Hubble parameter from $H=\frac{\dot{a}_{0}}{a_{0}}$.
Then equation (\ref{pap.cosm}) becomes
\begin{equation}\label{pap.hcosm}
2 H^{2} + \dot{H}= - \frac{k^{4}_{(5)}}{36} \rho(\rho+3p) -
k^{2}_{(5)} \frac{T_{55}}{3b^{2}_{0}}
\end{equation}
If one compares equation (\ref{pap.hcosm}) with usual Friedmann
equation, one can see that energy density enters the equation
quadratically, in contrast with the usual linear dependence.
Another novel feature of equation (\ref{pap.hcosm}) is that the
cosmological evolution depends on the five-dimensional Newton's
constant and not on the brane Newton's constant.

To have a feeling of what kind of cosmological evolution we get,
we consider the Bianchi identity $\nabla_{M}G^{M}_{~N}=0$. Then
using the Einstein equation (\ref{pap.beinst}) and the energy
momentum tensor on the brane from (\ref{pap.rpcom}), we get
\begin{equation}
\dot{\rho}+ 3(\rho+p) \frac{\dot{a}_{0}}{a_{0}}=0
\end{equation}
which is the usual energy density conservation. If we take for the
equation of state $p=w \rho$, then the above equation gives for
the energy density the usual relation
\begin{equation}
\rho \prec a_{0}  e^{-3(1+w)}
\end{equation}
If we look for power law solutions however, $a_{0}(t) \prec t^{q}$
equation (\ref{pap.hcosm}) gives
\begin{equation}
q=\frac{1}{3(1+w)}
\end{equation}
which comparing with $ q_{standard}=\frac{2}{3(1+w)}$ gives slower
expansion.

If we add a cosmological constant \index{cosmological constant} in
the bulk, then a solution of the Einstein equation
(\ref{pap.beinst}) can be obtained, in which the universe starts
with a non conventional phase and then enters the standard
cosmological phase \cite{pap.secbinet,pap.Cosm,pap.Stat}.

\section{Induced Gravity on the Brane}

The effective Einstein equations on the brane which we discussed
in section (3.1) were generalized in \cite{pap.maeda}, where
matter confined on the brane was taken under consideration.
 However, a more fundamental description of the physics
that produces the brane could include \cite{pap.sundrum} higher
order terms in a derivative expansion of the effective action,
such as a term for the scalar curvature of the brane, and higher
powers of curvature tensors on the brane. In \cite{pap.dvali1,
pap.dvali2} it was observed that the localized matter fields on
the brane (which couple to bulk gravitons) can generate via
quantum loops a localized four-dimensional worldvolume kinetic
term for gravitons. That is to say, four-dimensional gravity is
induced from the bulk gravity to the brane worldvolume by the
matter fields confined to the brane. We will therefore include the
scalar curvature term in our action and we will discuss what is
the effect of this term to cosmology.

Our theory is described then by the full action
(\ref{pap.action}). Using relations (\ref{pap.const}) we define a
distance scale
 \begin{equation}
r_{c}\equiv\frac{\kappa_{5}^{2}}{\kappa_{4}^{2}}=\frac{M_{4}^{2}}
{M_{5}^{3}}\,.
 \label{pap.distancescale}
 \end{equation}
Varying (\ref{pap.action}) with respect to the bulk metric
$g_{MN}$, we obtain the equations
\begin{equation}
^{(5)}G_{MN}=-\Lambda_{5}g_{MN}+\kappa_{5}^{2}\,(^{(5)}T_{MN}+\,^{(loc)}T_{MN}\,\delta(y))\,,
\label{pap.varying}
 \end{equation}
 where
\begin{equation}
^{(loc)}T_{MN}\equiv-\frac{1}{\kappa_{4}^{2}}\,\sqrt{\frac{^{(4)}g}
{^{(5)}g}}\,\,\Big{(} ^{(4)}G_{AB}-\kappa_{4}^{2}\,^{(4)}T_{MN}+
\Lambda_{4}g_{MN}\Big{)}
 \label{pap.tlocal}
\end{equation}
 is the localized energy-momentum tensor of the brane.
$^{(5)}G_{MN}$, $^{(4)}G_{MN}$ denote the Einstein tensors
constructed from the bulk and the brane metrics respectively.
Clearly, $^{(4)}G_{MN}$ acts as an additional source term for the
brane through $^{(loc)}T_{MN}$.

It is obvious that the additional source term on the brane will
modify the Israel Boundary Conditions (\ref{pap.isrmat}). The
modified conditions are
\begin{equation}
[K_{~\nu}^{\mu}]=-\kappa_{5}^{2}\,b_{o}\,\left(^{(loc)}T_{~\nu}^{\mu}-
\frac{^{(loc)}T}{3}\delta_{\nu}^{\mu}\right)\,, \label{pap.israel}
\end{equation}
 where the bracket means discontinuity of the extrinsic
curvature $K_{\mu\nu}=\frac{1}{2b}\partial_{y}g_{\mu\nu}$ across
$y=0$, and $b_{o}=b(y=0)$. A $\mathbf{Z}_{2}$ symmetry on
reflection around the brane is understood throughout.

Using equations (\ref{pap.varying}) and (\ref{pap.israel}) we can
derive the four-dimensional Einstein equations on the brane
\cite{pap.kofinas}. They are
\begin{equation}\label{pap.indequat}
^{(4)}G^{\mu}_{~\nu}= k^{2}_{(4)}~^{(4)}T^{\mu}_{~\nu}
-\Lambda_{(4)} \delta^{\mu}_{~\nu}- a \Big{(}L^{\mu}_{~\nu}
+\frac{L}{2} +\frac{3}{2} a \Big{)} \delta^{\mu}_{~\nu}
\end{equation}
where $\alpha\equiv 2/r_{c}$, while the quantities
$L^{\mu}_{~\nu}$ are related to the matter content of the theory
through the equation
 \begin{equation}
L_{\lambda}^{\mu} L_{\nu}^{\lambda}-\frac{L^{2}}{4} \,
\delta_{\nu}^{\mu} = \widetilde{T}_{\nu}^{\mu} - \frac{1}{4}
\Big{(} 3 \alpha^{2} + 2 \widetilde{T}_{\lambda}^{\lambda} \Big{)}
\delta_{\nu}^{\mu} \,, \label{pap.lll}
 \end{equation}
  and $L\equiv
L^{\mu}_{\mu}$. The quantities $\widetilde{T}^{\mu}_{\nu}$ are
given by the expression
\begin{eqnarray}
 \label{pap.energy}
\widetilde{T}_{\nu}^{\mu}&=&\Big(\Lambda_{4}-\frac{1}{2}\,
\Lambda_{5}\Big)\delta_{\nu}^{\mu}
-\kappa_{4}^{2}\,^{(4)}T_{\nu}^{\mu}+ \\&&
+\frac{2}{3}\,\kappa_{5}^{2}\,\Big(\,^{(5)}\overline{T}\,_{\nu}^{\mu}
+\Big(\,^{(5)}\overline{T}\,_{y}^{y}-\frac{^{(5)}\overline{T}}{4}\Big)\,\delta_{\nu}^{\mu}\Big)
-\overline{\textsf{E}}^{\,\mu}_{\,\nu}\,.
 \end{eqnarray}
Bars over $^{(5)}T^{\mu}_{\nu}$ and the electric part\,
$\textsf{E}^{\,^{\mu}}_{\,\nu}=C^{\mu}_{M \nu N}n^{M}n^{N}$ of the
Weyl tensor $C^{M}_{N P R}$ mean that the quantities are evaluated
at $y=0$. $\overline{\textsf{E}}^{\,\mu}_{\,\nu}$ carries the
influence of non-local gravitational degrees of freedom in the
bulk onto the brane  and makes the brane equations
(\ref{pap.indequat}) not to be, in general, closed. This means
that there are bulk degrees of freedom which cannot be predicted
from data available on the brane. One needs to solve the field
equations in the bulk in order to determine
$\textsf{E}^{\,^{\mu}}_{\,\nu}$ on the brane.

\subsection{Cosmology on the Brane with a $~ ^{(4)}R $ Term }

To get a feeling of what kind of cosmology we get on the brane
with an $ ^{(4)}R $ term, we consider the metric of
(\ref{pap.tmetric}). To simplify things we take $ ^{(5)}T_{MN} $
to be just the five-dimensional cosmological constant, while for
the matter localized on the brane we take $^{(4)}T_{MN}$ to have
the usual form of a perfect fluid (relation(\ref{pap.rpcom})). The
new term that enters here in the calculations is the
four-dimensional Einstein tensor $~ ^{(4)} G_{MN}$ which appears
in $~^{(loc)}T $ (relation (\ref{pap.tlocal})). The non vanishing
components of $~ ^{(4)}G_{MN}$ can be calculated to be

 \begin{eqnarray} \label{pap.rcomp}~~
 ~^{(4)}G_{00} &=&
-\frac{3 \delta (y)}{k^{2}_{4}b } \Big{\{} \frac{\dot{a}
^{2}}{a^{2}} + \frac{n^{2}}{a^{2}} \Big{\}} \nonumber  \\
^{(4)}G_{ij} &=& - \frac{\delta(y)}{k^{2}_{4}b } \Big{\{}
\frac{a^{2}}{n^{2}} \delta_{ij} \Big{(} - \frac{\dot{a}
^{2}}{a^{2}} + 2 \frac{\dot{a} \dot{n}}{an} - 2 \frac{
\ddot{a}}{a} \Big{)} -\delta _{ij} \Big{\}} \end{eqnarray}

 Then as in
section (3.1) we can calculate the distributional parts of the
second derivatives as \cite{deffayet}

\begin{eqnarray}
 \label{pap.rdistr} \frac{ [ a^{'} ] }{a_{0}b_{0}} &=&
-\frac{k^{2}}{3} \rho + \frac{k^{2}_{5}}{k^{2}_{4}n^{2}_{0}}
\Big{\{} \frac{\dot{a_{0}} ^{2}}{a^{2}_{0}} +
\frac{n^{2}_{0}}{a^{2}_{0}} \Big{\}} \nonumber
\\ \frac{ [ n^{'} ]}
{n_{0}b_{0}} &=& \frac{k^{2}}{3} (3 p + 2 \rho)
+\frac{k^{2}_{5}}{k^{2}_{4}n^{2}_{0}} \Big{(} - \frac{\dot{a}
^{2}_{0}}{a^{2}_{0}} - 2 \frac{\dot{a}_{0}
\dot{n}_{0}}{a_{0}n_{0}} + 2 \frac{ \ddot{a}_{0}}{a_{0}}  -
\frac{n^{2}_{0}}{a^{2}_{0}} \Big{)}\nonumber \\
\end{eqnarray}

If we again use a reflection symmetry $y\longrightarrow -y $, then
the Einstein equations (\ref{pap.varying}) with the use of
(\ref{pap.rpcom}), (\ref{pap.rcomp}) and (\ref{pap.rdistr}) give
our cosmological Einstein equation
\begin{equation}
\label{pap.efried}H^{2}- 2 \frac{k^{2}_{4}}{k^{2}_{5}} \sqrt{H^{2}
+ \frac{1}{a^{2}_{0}}} = - \frac{k^{2}}{3} \rho +
\frac{1}{a^{2}_{0}}
 \end{equation}

To compare this equation with the evolution equation
(\ref{pap.cosm}) we had derived without the $~^{(4)}R $ term, we
observe that the energy density enters the evolution equation
linearly as in the standard cosmology. However the evolution
equation (\ref{pap.efried}) is not the standard Friedmann
cosmological equation. We can recover the usual Friedmann equation
\cite{dick}, if (neglecting the $\frac{1}{a^{2}_{0}}$ term)
 \begin{equation}
H^{-1}\ll \frac{M^{2}_{(4)}}{2M^{3}_{(5)}}
 \end{equation}
 If we use the crossover scale $r_{c}$ of
 (\ref{pap.distancescale}) the above relation means that an
observer on the brane will see correct Newtonian gravity at Hubble
distances shorter than a certain crossover scale, despite the fact
that gravity propagates in extra space which was assumed there to
be flat with infinite extent; at larger distances, the force
becomes higher-dimensional. We can get the same picture if we look
at the equation (\ref{pap.lll}). If the crossover scale $r_{c}$ is
large, then $\alpha\equiv 2/r_{c}$ is small and the last term in
(\ref{pap.lll}) decouples, giving the usual four-dimensional
Einstein equations.

\section{A brane on a Move}

So far the domain walls (branes) were static solutions of
 the underlying theory, and the cosmological evolution of our
 universe was due mainly to the time evolution of energy density on the
 domain wall (brane). In this section we will consider
  another  approach. The
 cosmological evolution of our universe is due to the motion of
 our brane-world in the background gravitational field of the
 bulk \cite{pap.Cha,pap.kraus,pap.Keh,pap.papantono}.

 In \cite{pap.Cha} the motion of a domain wall (brane) in a higher
 dimensional spacetime was studied. The Israel matching conditions
 were used to relate the bulk to the domain wall (brane) metric, and
 some interesting cosmological solutions were found. In
 \cite{pap.Keh} a universe three-brane is considered in motion
 in ten-dimensional space in the presence of a gravitational field
 of other branes. It was shown that this motion in ambient space
 induces cosmological expansion (or contraction) on our universe,
 simulating various kinds of matter. In particular, a D-brane  moving
 in a generic
 static, spherically symmetric background was considered. As the brane moves in a
 geodesic, the induced world-volume metric becomes a function of
 time, so there is a cosmological evolution from the brane point
 of view. The metric of a three-dimensional brane is parametrized as

\begin{equation}\label{pap.inmet}
ds^{2}_{10}=g_{00}(r)dt^{2}+g(r)(d\vec{x})^{2}+
  g_{rr}(r)dr^{2}+g_{S}(r)d\Omega_{5}
\end{equation}
 and there is also a dilaton field $\Phi$ as well as a $RR$
 background~$C(r)=C_{0...3}(r)$ with a self-dual field strength. The
 action on the brane is given by

\begin{eqnarray}\label{pap.BIaction}
  S&=&T_{3}~\int~d^{4}\xi
  e^{-\Phi}\sqrt{-det(\hat{G}_{\mu\nu}+(2\pi\alpha')F_{\mu\nu}-
  B_{\mu\nu})}
   \nonumber \\&&
  +T_{3}~\int~d^{4}\xi\hat{C}_{4}+anomaly~terms
\end{eqnarray}
 The induced metric on the brane is
\begin{equation}\label{pap.indmetric}
  \hat{G}_{\mu\nu}=G_{MN}\frac{\partial x^{M}\partial x^{N}}
  {\partial\xi^{\mu}\partial\xi^{\nu}}
\end{equation}
 with similar expressions for $F_{\alpha\beta}$ and
 $B_{\alpha\beta}$. In the static gauge,
 using (\ref{pap.indmetric}) we can calculate the bosonic part of the
 brane Lagrangian which reads

\begin{equation}\label{pap.braneLagr}
L=\sqrt{A(r)-B(r)\dot{r}^{2}-D(r)h_{ij}\dot{\varphi}^{i}\dot{\varphi}^{j}}
-C(r)
\end{equation}
where $h_{ij}d \varphi ^{i} d \varphi^{j}$ is the line
 element of the unit five-sphere, and

\begin{equation}\label{pap.metfun}
  A(r)=g^{3}(r)|g_{00}(r)|e^{-2\Phi},
  B(r)=g^{3}(r)g_{rr}(r)e^{-2\Phi},
  D(r)=g^{3}(r)g_{S}(r)e^{-2\Phi}
\end{equation}

 Demanding conservation of energy $E$ and of total angular
 momentum $ \ell ^{2} $ on the brane, the induced four-dimensional metric
 on the brane is

\begin{equation}\label{pap.fmet}
d\hat{s}^{2}=(g_{00}+g_{rr}\dot{r}^{2}+g_{S}h_{ij}\dot{\varphi}^{i}\dot{\varphi}^{j})dt^{2}
+g(d\vec{x})^{2}
\end{equation}
 with

\begin{equation}\label{pap.functions}
\dot{r}^{2}=\frac{A}{B}(1-\frac{A}{(C+E)^{2}}\frac{D+\ell^{2}}{D}),
h_{ij}\dot{\varphi}^{i}\dot{\varphi}^{j}=\frac{A^{2}\ell^{2}}{D^{2}(C+E)^{2}}
\end{equation}
 Using (\ref{pap.functions}), the induced metric becomes

\begin{equation}\label{pap.finindmetric}
d\hat{s}^{2}=-d\eta^{2}+g(r(\eta))(d\vec{x})^{2}
\end{equation}
 with $\eta$ the cosmic time which is defined  by
\begin{equation}\label{pap.cosmic}
 d\eta=\frac{|g_{00}|g^{\frac{3}{2}}e^{-\Phi}}{|C+E|}dt
\end{equation}

 This equation is the standard form of a flat expanding universe.
If we define the scale factor as $\alpha^{2}=g$ then we can
calculate the Hubble constant $H=\frac{\dot{\alpha}}{\alpha}$,
where dot stands for derivative with respect to cosmic time. Then
we can interpret the quantity $(\frac{\dot{\alpha}}{\alpha})^{2}$
as an effective matter density on the brane with the result
\begin{equation}\label{pap.dens}
\frac{8\pi}{3}\rho_{eff}=\frac{(C+E)^{2}g_{S}e^{2\Phi}-|g_{00}|(g_{S}g^{3}+\ell^{2}e^{2\Phi})}
{4|g_{00}|g_{rr}g_{S}g^{3}}(\frac{g'}{g})^{2}
\end{equation}

Therefore the motion of a three-dimensional brane on a general
spherically symmetric background had induced on the brane a matter
density. As it is obvious from the above relation, the specific
form of the background will determine the cosmological evolution
on the brane.

We will go to a particular background, that of a Type-0 string,
and see what cosmology we get. The action of the Type-0 string is
given by \cite{pap.Tset}
\begin{eqnarray}\label{pap.straction}
S_{10}&=&~\int~d^{10}x\sqrt{-g}\Big{[} e^{-\Phi} \Big{(}
 R+(\partial_{\mu}\Phi)^{2} -\frac{1}{4}(\partial_{\mu}T)^{2}
-\frac{1}{4}m^{2}T^{2}-\frac{1}{12}H_{mnr}H^{mnr}\Big{)} \nonumber
\\&& - \frac{1}{2}(1+T+\frac{T^{2}}{2})|F_{5}|^{2} \Big{]}
\end{eqnarray}

The equations of motion which result from this action are
\begin{equation}\label{pap.dilaton}
  2\nabla^{2}\Phi-4(\nabla_{n}\Phi)^{2}-\frac{1}{2}m^{2}T^{2}=0
\end{equation}

\begin{eqnarray}\label{pap.metric}
  &R_{mn}&+2\nabla_{m}\nabla_{n}\Phi-\frac{1}{4}\nabla_{m}T\nabla_{n}T
  -\frac{1}{4\cdot4!}e^{2\Phi}f(T) \Big{(}F_{mklpq}F_{n}~^{klpq}\nonumber
  \\&&  - \frac{1}{10}G_{mn}F_{sklpq}F^{sklpq} \Big{)}=0
\end{eqnarray}

\begin{equation}\label{pap.Tachyon}
  (-\nabla^{2}+2\nabla^{n}\Phi\nabla_{n}+m^{2})T
  +\frac{1}{2\cdot5!}e^{2\Phi}f'(T)F_{sklpq}F^{sklpq}=0
\end{equation}
\begin{equation}\label{pap.F}
  \nabla_{m} \Big{(}f(T)F^{mnkpq} \Big{)}=0
\end{equation}
The tachyon is coupled to the $RR$ field through the function
\begin{equation}\label{pap.ftac}
 f(T)=1+T+\frac{1}{2} T^{2}
\end{equation}
In the background where the tachyon field acquires vacuum
expectation value $T_{vac}=-1$, the tachyon function
(\ref{pap.ftac}) takes the value $f(T_{vac})=\frac{1}{2}$ which
guarantee the stability of the theory \cite{pap.Kleb}.

The equations (\ref{pap.dilaton})-(\ref{pap.F}) can be solved
using the metric (\ref{pap.inmet}). Moreover one can find the
electrically charged three-brane if
 the following
ansatz for the $RR$ field

\begin{equation}\label{pap.Form}
  C_{0123}=A(r),   F_{0123r}=A'(r)
\end{equation}
and a constant value for the dilaton field $\Phi=\Phi_{0}$ is used

\begin{equation}\label{pap.Sol}
  g_{00}=-H^{-\frac{1}{2}},
  g(r)=H^{-\frac{1}{2}},  g_{S}(r)=H^{\frac{1}{2}}r^{2},
g_{rr}(r)=H^{\frac{1}{2}},    H=1+\frac{e^{\Phi_{0}}Q}{2r^{4}}
\end{equation}

\subsection{Cosmology of the Moving Brane}

 The induced metric on the brane (\ref{pap.fmet})
using the background solution (\ref{pap.Sol}) is
\begin{equation}\label{pap.IndSol}
 d\hat{s}^{2}=(-H^{-\frac{1}{2}}+H^{\frac{1}{2}}\dot{r}^{2}
 +H^{\frac{1}{2}}r^{2}h_{ij}\dot{\varphi}^{i}\dot{\varphi}^{j})dt^{2}
 +H^{-\frac{1}{2}}(d\vec{x})^{2}
\end{equation}
From equation (\ref{pap.F}) the $RR$ field $C=C_{0123}$ using the
ansatz (\ref{pap.Form}) becomes
\begin{equation}\label{pap.cbar}
  C^{~'}=2 Q g^{2}g^ {-\frac{5}{2}}_{s}\sqrt{g_{rr}}f^{-1}(T)
\end{equation}
where Q is a constant. Using again the solution (\ref{pap.Sol})
the $RR$ field can be integrated to give

\begin{equation}\label{pap.Cterm}
C=e^{-\Phi_{0}}f^{-1}(T)(1+\frac{e^{\Phi_{0}}Q}{2r^{4}})^{-1}+Q_{1}
\end{equation}
where $Q_{1}$ is a constant. The effective density on the brane
(\ref{pap.dens}), using equation (\ref{pap.Sol}) and
(\ref{pap.cbar}) becomes \cite{pap.papa}
\begin{equation}\label{pap.cre}
\frac{8\pi}{3}\rho_{eff}=\frac{1}{4}[(f^{-1}(T)+EHe^{\Phi_{0}})^{2}-(1+\frac{\ell^{2}e^{
2\Phi_{0}}}{2}H)]
\frac{Q^{2}e^{2\Phi_{0}}}{r^{10}}H^{-\frac{5}{2}}
\end{equation}
where the constant $Q_{1}$ was absorbed in a redefinition of the
parameter $E$. Identifying $g=\alpha^{2}$ and using
$g=H^{-\frac{1}{2}}$ we get from (\ref{pap.cre})
\begin{eqnarray}\label{pap.aro}
\frac{8\pi}{3}\rho_{eff}&=&(\frac{2e^{-\Phi_{0}}}{Q})^{\frac{1}{2}}
 \Big{[} \Big{(} f^{-1}(T)+\frac{Ee^{\Phi_{0}}}{\alpha^{4}} \Big{)}^{2}
-\Big{(}1+\frac{\ell^{2}e^{2\Phi_{0}}}
{\alpha^{6}}(\frac{2e^{-\Phi_{0}}}{Q})^{\frac{1}{2}}\nonumber \\&&
(1-\alpha^{4})^{\frac{1}{2}} \Big{)}  \Big{]} (1-\alpha^{4})
^{\frac{5}{2}}
\end{eqnarray}
From the relation $g=H^{-\frac{1}{2}}$ we find
\begin{equation}\label{pap.ro}
  r= (\frac{\alpha^{4}}{1-\alpha^{4}})
  ^{\frac{1}{4}}(\frac{Qe^{\Phi_{0}}}{2})^{\frac{1}{4}}
\end{equation}
This relation restricts the range of $\alpha$ to $0\leq \alpha
<1$, while the range of $r$ is $0\leq r< \infty$. We can calculate
the scalar curvature of the four-dimensional universe as
\begin{equation}\label{pap.curv}
  R_{brane}=8\pi(4+\alpha\partial_{\alpha})\rho_{eff}
\end{equation}
If we use the effective density of (\ref{pap.aro}) it is easy to
see that $R_{brane}$ of (\ref{pap.curv}) blows up at $\alpha=0$.
On the contrary if $r\rightarrow 0$, then the $ds^{2}_{10}$ of
(\ref{pap.inmet}) becomes
\begin{equation}\label{pap.ads}
ds^{2}_{10}= \frac{r^{2}}{L} (-dt^{2}+(d\vec{x})^{2})+
      \frac{L}{r^{2}} dr^{2}+  L d\Omega_{5}
\end{equation}
with $L=(\frac{e^{\Phi_{0}}Q}{2})^{\frac{1}{2}}$. This space  is a
regular $AdS_{5} \times S^{5}$ \index{Anti de-Sitter space} space.

Therefore the brane develops an initial singularity as it reaches
$r=0$, which is a coordinate singularity and otherwise a regular
point of the $AdS_{5}$ space. This is another example in Mirage
Cosmology \cite{pap.Keh} \index{mirage cosmology} where we can
understand the initial singularity as the point where the
description of our theory breaks down.

If we take $\ell^{2}=0$, set the function $f(T)$ to each minimum
value and also taking $\Phi_{0}=0$, the effective density
(\ref{pap.aro}) becomes
\begin{equation}\label{pap.laro}
\frac{8\pi}{3}\rho_{eff}=(\frac{2}{Q})^{\frac{1}{2}}
\Big{(}(2+\frac{E}{\alpha^{4}})^{2} -1 \Big{)} (1-\alpha^{4}) ^{\frac{5}{2}}
\end{equation}
As we can see in the above relation, there is a constant term,
coming from the tachyon function $f(T)$. For small $\alpha$ and
for some range of the parameters $E$ and $Q$ it gives an
inflationary phase to the brane cosmological evolution. In Fig.5
we have plotted $\rho_{eff}$ as a function of $\alpha$ for $Q=2$.
 Note here that $E$ is constrained from
(\ref{pap.functions}) as $C+E\geq0$. In our case using
(\ref{pap.Cterm}) we get $E\geq -2\alpha^{4}$, therefore $E$ can
be as small as we want.

\begin{figure}[h]
\centering
\includegraphics[scale=0.7]{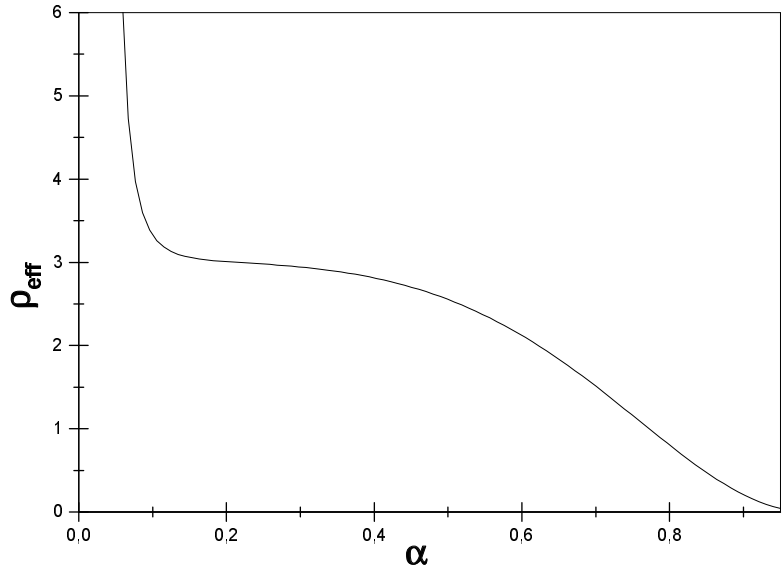}
\caption {The induced energy density on the brane as a function of
the brane scale factor.}
\end{figure}

The cosmological evolution of a brane universe according to this
example is as follows. As the brane moves away from $r=0$ to
larger values of $r$, the universe after the inflationary phase
enters a radiation dominated epoch because the term $\alpha^{-4}$
takes over in (\ref{pap.laro}). As the cosmic time $\eta$ elapses
the $\alpha^{-8}$ term dominates and finally when the brane is far
away from $r=0$, the term which is controlled by the angular
momentum $\ell^{2}$ gives the main contribution to the effective
density. Non zero values of $\ell^{2}$ will give negative values
for $\rho_{eff}$. We expect that at later cosmic times there will
be other fields, like gauge fields, which will give a different
dynamics to the cosmological evolution and eventually cancel the
negative matter density.

The above model can be generalized to include a non constant value
for the dilaton field. Then using \cite{pap.Typ0} and
\cite{pap.Minah} we can study the cosmological evolution of a
brane universe as the brane moves from IR to UV in the background
of a type-0 string theory
\cite{pap.pappa,pap.kim,pap.korean,pap.youm}.

\section{Conclusions}

We presented the main ideas and gave the main results of the
cosmological evolution of a brane universe. The main new result
that brane cosmology offered, is that our universe at some stage
of its evolution, passed a cosmological phase which is not
described purely by the Friedmann equation of standard cosmology.
In the simplest possible brane model, the Hubble parameter scales
like the square of the energy density and this results in a slower
universe expansion. There were a lot of extensions and
modifications of this model, trying to get the standard cosmology
but it seems that the universe in a brane world passed from an
unconvensional phase at its earliest stages of its cosmological
evolution.

The inclusion of an $~^{(4)}R $ term in the action, offered a more
natural explanation of the brane unconvensional phase. At small
cosmological distances our universe was involved according the
usual Einstein equations. If the cosmological scale is larger than
a crossover scale, we enter a higher-dimensional regime where the
cosmological evolution of our brane universe is no longer coverned
by the conventional Friedmann equation.

We also presented a model where a brane is moving in the
gravitational field of other branes. Then we can have the standard
cosmological evolution on the brane, with the price to be paid,
that the matter on the brane is a "mirage" matter.

 \end{document}